\documentclass[12pt]{article}
\usepackage{fullpage}
\usepackage{hyperref}
\hypersetup{colorlinks=true}

\usepackage{threeparttable}
\usepackage{lscape}

\usepackage{graphicx}
\usepackage{caption}
\usepackage{subcaption}

\usepackage{amsmath}
\usepackage{longtable}
\usepackage[none]{hyphenat}
 \usepackage{setspace}
 \usepackage{url}

\usepackage{lineno}
\usepackage{epstopdf}
\usepackage{caption}

\newcommand{\Emax}{\emph{E. maximus indicus}}

\newcommand{\Mprim}{\emph{M. primigenius}}
\newcommand{\MprimFull}{\emph{Mammuthus primigenius}}
\newcommand{\Lafr}{\emph{L. africana}}
\newcommand{\LafrFull}{\emph{Loxodonta africana}}

\newcommand{\Rebekah}{Rebek\mbox{}ah }
\newcommand{\Fox}{\emph{FOXQ1}}

\makeatletter
\renewcommand{\@biblabel}[1]{\quad#1.}
\makeatother
\usepackage{natbib}
\bibpunct[; ]{(}{)}{,}{a}{}{;}

\usepackage{textcomp}

\begin{document}

\author{\Rebekah L. Rogers$^1$ and Montgomery Slatkin$^1$}

\title{Excess of genomic defects in a woolly mammoth on Wrangel island}
\date{}

\maketitle

\begin{center} \Large Research Article \end{center}
\vspace{0.25in}

\noindent 1) Dept of Integrative Biology, University of California, Berkeley \\

\vspace{0.25in}

\noindent \textbf{Running head: }Mutational meltdown in woolly mammoths

\vspace{0.25in}


\vspace{0.25in}

\noindent \textbf{Corresponding author:} Rebekah L. Rogers,  Dept. of Integrative Biology University of California, Berkeley, CA 94720 \\
\\
\noindent \textbf{Phone:}  949-824-0614

\noindent \textbf{Fax:}  949-824-2181

\noindent \textbf{Email:} bekah@berkeley.edu

\newpage


\section*{Abstract}
Woolly mammoths (\MprimFull) populated Siberia, Beringia, and North America during the Pleistocene and early Holocene.  Recent breakthroughs in ancient DNA sequencing have allowed for complete genome sequencing for two specimens of woolly mammoths (Palkopoulou et al. 2015).  One mammoth specimen is from a mainland population ~45,000 years ago when mammoths were plentiful.   The second, a  4300 yr old specimen, is derived from an isolated population on Wrangel island where mammoths subsisted with small effective population size more than 43-fold lower than previous populations.  These extreme differences in effective population size offer a rare opportunity to test nearly neutral models of genome architecture evolution within a single species.   Using these previously published mammoth sequences, we identify deletions, retrogenes, and non-functionalizing point mutations.   In the Wrangel island mammoth, we identify a greater number of deletions, a larger proportion of deletions affecting gene sequences, a greater number of candidate retrogenes, and an increased number of premature stop codons.  This accumulation of detrimental mutations is consistent with genomic meltdown in response to low effective population sizes in the dwindling mammoth population on Wrangel island.   In addition, we observe high rates of loss of olfactory receptors and urinary proteins, either because these loci are non-essential or because they were favored by divergent selective pressures in island environments.  Finally, at the locus of \emph{FOXQ1} we observe two independent loss-of-function mutations, which would confer a satin coat phenotype in this island woolly mammoth.  

\clearpage{}

\section*{Author Summary}
We observe an excess of detrimental mutations, consistent with genomic meltdown in woolly mammoths on Wrangel Island just prior to extinction.  We observe an excess of deletions, an increase in the proportion of deletions affecting gene sequences, and an excess of premature stop codons in response to evolution under low effective population sizes.   Large numbers of olfactory receptors appear to  have loss of function mutations in the island mammoth.  These results offer genetic support within a single species for nearly-neutral theories of genome evolution.  We also observe two independent loss of function mutations at the \emph{FOXQ1} locus, likely conferring a satin coat in this unusual woolly mammoth.

\clearpage

\section*{Introduction}

Woolly mammoths (\MprimFull) were among the most populous large herbivores in North America, Siberia, and Beringia during the Pleistocene and early Holocene \citep{Stuart2004}.  However warming climates and human predation led to extinction on the mainland roughly 10,000 years ago \citep{Nogues2008}.  Lone isolated island populations persisted out of human reach until roughly 3,700 years ago when the species finally went extinct \citep{Vartanyan2008}.  Recently, two complete high-quality high-coverage genomes were produced for two woolly mammoths \citep{Palkopoulou2015}.  One specimen is derived from the Siberian mainland at Oimyakon, dated to 45,000 years ago \citep{Palkopoulou2015}.  This sample comes from a time when mammoth populations were plentiful, with estimated effective population size of $N_e=13,000$ individuals \citep{Palkopoulou2015}.  The second specimen is from Wrangel Island off the north Siberian coast \citep{Palkopoulou2015}.  This sample from 4,300 years ago represents one of the last known mammoth specimens.   This individual comes from a small population  estimated to contain roughly 300 individuals \citep{Palkopoulou2015}.  These two specimens offer the rare chance to explore the ways the genome responds to pre-extinction population dynamics.  

Nearly neutral theories of genome evolution predict that small population sizes will lead to an accumulation of detrimental variation in the genome \citep{LynchBook}.  Such explanations have previously been invoked to explain genome content and genome size differences across multiple species \citep{Lynch2006}.  Yet, within-species comparisons of how genomes are changed by small effective population sizes remain necessarily rare.  These mammoth specimens offer the unique opportunity for within-species comparative genomics under a 43-fold reduction in population size.  This comparison offers a major advantage as it will be free from confounding biological variables that are present in cross species comparisons.   If nearly neutral dynamics lead to an excess of detrimental variation, we should observe an excess of harmful mutations in pre-extinction mammoths from Wrangel Island.  

We use these two ancient DNA sequences to identify retrogenes, deletions, premature stop codons, and point mutations found in the Wrangel Island and Oimyakon mammoths.  We identify an excess of putatively detrimental mutations, with an excess of stop codons, an excess of deletions, an increase in the proportion of deletions affecting gene sequences, an increase in non-synonymous substitutions relative to synonymous substitutions, and an excess of retrogenes, reflecting increased transposable element activity.  These data bear the signature of genomic meltdown in small populations, consistent with nearly-neutral genome evolution.  They furthermore suggest large numbers of detrimental variants collecting in pre-extinction genomes, a warning for continued efforts to protect current endangered species with small population sizes.

\section*{Results}
\subsection*{Excess of amino acid substitutions and stop codons}
\begin{figure}
 \center
\includegraphics{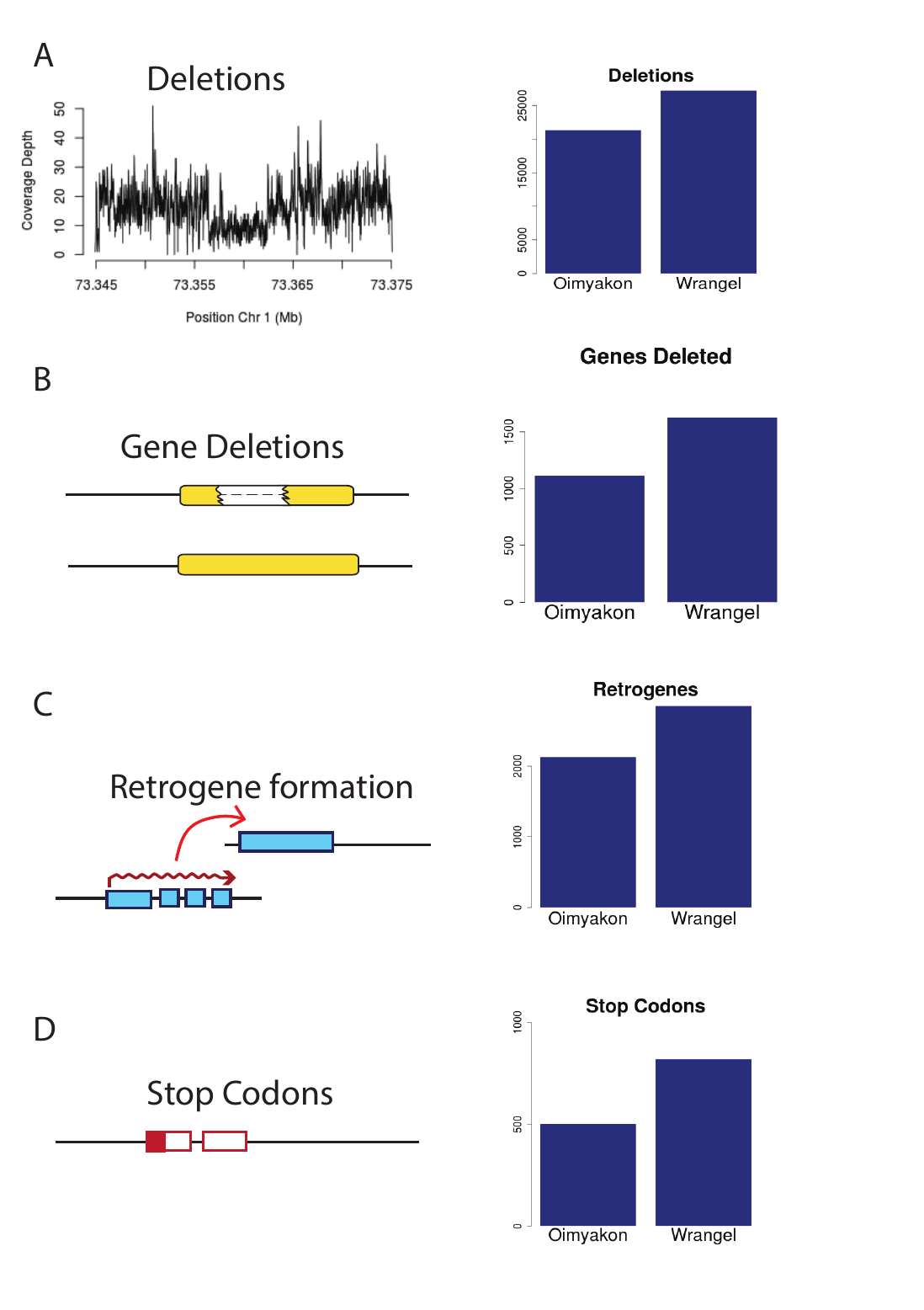}

\caption{\label{Fig1}  Excess of putatively detrimental mutations in the Wrangel Island Genome.  A) Deletions B)Genes deleted C) Retrogenes D) Premature stop codons.  Numbers shown are corrected for false negative rates of 30\% for heterozygous SNPs and 0.5\% for deletions in the lower coverage Oimyakon mammoth. }
\end{figure}

  We identified all SNPs in each mammoth genome as well as one Indian elephant specimen, Maya, using GATK \citep{GATK}.  We identified all non-synonymous and synonymous changes relative to the \Lafr {} reference genome (\url{https://www.broadinstitute.org/scientific-community/science/projects/mammals-models/elephant/elephant-genome-project}) using r3.7 annotations lifted over to \Lafr {} 4.0 genome sequences.  We observe a significant increase in the number of heterozygous non-synonymous changes relative to synonymous changes in the Wrangel island genome compared with Oimyakon ($\chi^2= 68.799$, $df = 1$, $P < 2.2\times10^{-16}$; Table \ref{ChiSqTable}).  There is also a significant increase in the number of homozygous mutations at non-synonymous sites  relative to synonymous sites ($\chi^2= 9.96$, $df = 1$, $P < 0.0016$; Table \ref{ChiSqTable}).  We further observe an excess of premature stop codons in the genome of the Wrangel Island mammoth, with ~1.8X as many genes affected.  There are 503 premature stop codons in the Oimyakon genome (adjusting for a 30\% false negative rate at heterozygous sites) compared with 819 in the Wrangel island genome (Figure \ref{Fig1}).  There are 318 genes that have premature stop codons that are shared across the two mammoths, and 357 genes that are truncated in both mammoths, including mutations that form at independent sites.    A total of 120 of these genes have stop codons in the two mammoths as well as in Maya the Indian elephant, suggesting read through in the \Lafr {} reference.  Among truncated genes, there is a significant excess of olfactory genes and oderant binding receptors that appear to be pseudogenized with an EASE enrichment score of 9.1 (Table \ref{GOStops}) \citep{DAVID1,DAVID2}.      We observe 85 truncated olfactory receptors and 3 vomeronasal receptors as well as multiple signal transduction peptides compared with 44 olfactory receptors and 2 vomeronasal receptors pseudogenized in the mainland mammoth.
  
It is possible that DNA damage in the archaic specimens could contribute to a portion of the observed stop codons.  When we exclude A/G and C/T mutations, there is still a gross excess of premature stop codons, with 645 genes truncated in the Wrangel Island mammoth compared with 377 in the Oimyakon mammoth.  Hence, the patterns are not explained solely by differential DNA damage in the two mammoths.  Maya, the Indian Elephant specimen shows 450 premature stop codons, but 401 when A/G and T/C mutations are excluded.  When putative damage to ancient DNA is excluded, Maya appears to house an intermediate number of premature stop codons, with a 6\% increase compared to the Oimyakon mammoth.     

\section*{Deletions}

We identify 27228 deletions over 1 kb long in the Wrangel island genome, and 21346 (correcting for a 0.5\% false negative rate at heterozygous sites) in the Oimyakon genome (Table \ref{MutTypes}).  There are 6147 deletions (23\%) identified in the Wrangel Island mammoth that are homozygous ($\leq 10\%$ coverage) compared with 5035 (24\%) in the Oimyakon mammoth.  (Table \ref{HetDels}). A total of 13,459 deletions are identified in both mammoth genomes (Table \ref{Shared}).   Some 4813 deletions in the Wrangel Island mammoth  and 4598 in the Oimyakon mammoth appear hemizygous but have stretches of zero coverage for at least 50\% of their length.  These sites may represent multiple independent homozygous deletions that cannot be differentiated via change point statistics.  Alternatively, they might indicate smaller secondary deletions that appear on hemizygous haplotypes.  Such secondary deletions are common when large loop mismatch repair attacks unpaired, hemizygous stretches of DNA \citep{Rogers2014, kearney2001}.  The Wrangel Island Mammoth has sharply increased heterozygosity for deletions in comparison with the Oimyakon mammoth (Table \ref{HetDels}).  Some portion of the inflated heterozygosity for deletions in the Wrangel Island mammoth could be due to this difficulty in inferring genotypes in a high throughput setting.   Alternatively, the effective mutation rate may have increased as fewer deletions were removed from the population via purifying selection, inflating $\theta_{del}$.  It is also possible that there was an increase in the rate of deletions in the Wrangel Island lineage due to defective DNA repair mechanisms.   An increase in non-homologous end joining after DNA breaks rather than double stranded break repair could putatively induce such a change in the deletion rate.   

Maya the Indian elephant shows a larger number of deletions than the Oimyakon mammoth, but with different character from the Wrangel Island mammoth. The bulk of these are derived from 22,954 hemizygous deletions (Table \ref{HetDels}).  Maya houses only 5141 homozygous deletions, similar to the mainland mammoth (Table \ref{HetDels}).  There is an increase in the number of hemizygous deletions that affect gene sequences, but only a modest increase in the number of homozygous deletions that affect gene sequences (Table \ref{HetDels}).   Competing pressures of higher $N_e$, longer time frames to accumulate mutations toward equilibrium frequencies, differences in mutation rates between the mammoths and elephants, differences in selective pressures, differences in  the distribution of selective coefficients for deletions, different effective mutation rates due to different selective constraints, or differences in dominance coefficients might all contribute to differences in the number of deletions observed in elephants and mammoths.  Additional samples would be necessary to determine the extent to which genetic declines may be influencing the diversity of deletions in modern Indian elephants.  We currently have no basis for conclusions given this single sample, with no prior comparison.

\begin{figure}
 \center
\includegraphics[scale=0.5]{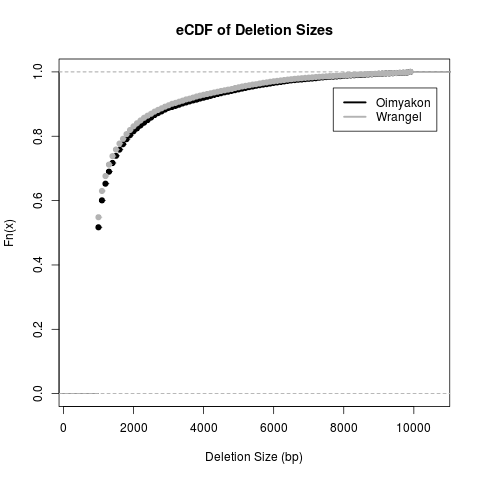}
\caption{\label{SizeFig}  eCDF for the size distribution of deletions in the Oimyakon and Wrangel Island genomes.  There is a significant reduction in the size of deletions identified in the Wrangel Island Genome. }
\end{figure}

There is a significant difference in the size distribution of deletions identified in the two mammoth samples, with a mean of 1707 bp in Oimyakon and 1606 bp in the Wrangel mammoth (Wilcox $W = 304430000$, $P < 2.2e-16$; Figure \ref{SizeFig}).  This difference could reflect either differences in DNA replication or repair mechanisms in the two mammoths, or altered selective constraints for different types of duplications. No significant difference is observed between the Wrangel island mammoth down sampled sequence data   ($W = 2004400$, $P= 0.3917$) suggesting that the observed decrease in size is not due to differences in coverage.     Some 1628 genes have deleted exons in the Wrangel Island mammoth compared to 1110 in Oimyakon (Table \ref{MutTypes}), a significant excess of genes deleted compared to expectations based on the number of deletions ($\chi^2 = 12.717$, $df = 1$,$P= 0.0003623$).  Among these deleted genes, 112 in the mainland mammoth are homozygous compared to 133 homozygous exon deletions in the Wrangel Island Mammoth.   Gene functions for affected genes in the Oimyakon mammoth include synapse functions, PHD domains,  zinc fingers, aldo-keto metabolism, calcium dependent membrane targeting, DNA repair, transcription regulation, and development (Table \ref{GODels}).   Gene functions overrepresented among deletions in the Wrangel Island mammoth include major urinary proteins, lipocalins, and pheromones, pleckstrins, transcription regulation, cell transport, DNA repair, chromatin regulation, hox domains, and  development  (Table \ref{GODels}).

Among the genes deleted in the Wrangel Island mammoth, several have phenotypes of interest in other organisms.  We observe a hemizygous deletion in riboflavin kinase \emph{RFK} in the Wrangel Island mammoth, but normal coverage in the Oimyakon mainland mammoth (Figure \ref{RFKDepth}).  Homozygous knockouts of riboflavin kinase,  essential for B2 utilization/FAD synthesis, are embryonic lethal in mice \citep{yazdanpanah2009}.  Finally,  we identify a hemizygous deletion in the Wrangel island mammoth that would remove the entire gene sequence at the \Fox {} locus (Figure \ref{SatinDepth}).   The alternative haplotype carries a frameshift mutation that disrupts the FOXQ1 functional domain.  \Fox {} knock-outs in mice are associated with the satin coat phenotype, which results in translucent fur but normal pigmentation due to abnormal development of the inner medulla of hairs \citep{Hong2001}, with two independent mutations producing this phenotype \citep{Hong2001}.   \Fox {} also regulates mucin secretion in the GI tract,  a case of pleiotropic functions from a single gene \citep{Verzi2008}.  If the phenotype in elephantids matches the phenotype exhibited in mice, this mammoth would have translucent hairs and a shiny satin coat, caused by two independently formed knock-out alleles at the same locus.  These genes each have functions that are conserved across mammals, though there is no guarantee that they would produce identical phenotypes in other species.  

\subsection*{Retrogene formation}
Retrogene formation can serve as a proxy for retrotransposon activity.  We identify retrogenes that display exon-exon junction reads in genomic DNA.   We observe 1.3X more retrogenes formed in the Wrangel island mammoth.  The Wrangel Island mammoth has 2853 candidate retrogenes, in comparison with 2130 in the Oimyakon mammoth  and 1575 in Maya (Table \ref{MutTypes}).  There are 436 retrogenes that are shared between the two mammoths, though some of these could arise via independent mutations. This excess of retrogenes is consistent with increased retroelement activity in the Wrangel Island lineage.    During retrogene formation, highly expressed genes, especially those expressed in the germline, are expected to contribute to new retrogenes.  To determine the types of loci that had been copied by retrotransposons, we performed a gene ontology analysis using DAVID \citep{DAVID1,DAVID2}.  Functional categories overrepresented among candidate retrogenes include genes involved in transcription, translation, cell division/cytoskeleton, post translational modification, ubiquitination, and chaperones for protein folding (Table \ref{GORetrosOimyakon}-\ref{GORetrosWrangel}).    All of these are expected to be highly expressed during cell divisions or constitutively  expressed, consistent with expectations that highly expressed genes will be overrepresented.  Gene ontologies represented are similar for both mammoths (Table \ref{GORetrosOimyakon}-\ref{GORetrosWrangel}). Although these retrogenes are unlikely to be detrimental in and of themselves, they may point to a burst of transposable element activity in the lineage that led to the Wrangel island individual.  Such a burst of TE activity would be expected to have detrimental consequences, additionally contributing to genomic decline.    
 
\subsection*{Genomic effects of demography}
Under nearly-neutral theory of genome evolution, detrimental mutations should accumulate in small populations as selection becomes less efficient \citep{LynchBook}.   This increase in non-neutral amino acid changes and premature stop codons is consistent with reduced efficacy of selection in small populations.  We attempted to determine whether the data is consistent with this nearly-neutral theory at silent and amino acid replacement substitutions whose mutation rates and selection coefficients are well estimated in the literature. Under nearly neutral theory, population level variation for non-synonymous amino acid changes should accelerate toward parity with population level variation at synonymous sites.  

Given the decreased population size on Wrangel Island, we expect to observe an accumulation of detrimental changes that would increase heterozygosity at non-synonymous sites ($H_N$) relative to synonymous sites ($H_S$) in the island mammoth.   Heterozygosity depends directly on effective population sizes.  We observe $H_S=0.00130\pm0.00002$ in the Wrangel Island mammoth, which is 80\% of $H_S=0.00161\pm0.00002$ observed in the Oimyakon mammoth (Table \ref{HetNS}).  The magnitude of the difference between $H_S$ in these two mammoths is 28 standard deviations apart, suggesting that these  two mammoths could not have come from populations with the same effective population sizes.  The specimens are well beyond the limits of expected segregating variation for a single population. To determine whether such results are consistent with theory, we fitted a model using PSMC inferred population sizes for the Wrangel island mammoth, based on decay of heterozygosity of $(1-1/2N)^t H_0$.   The observed reduction in heterozygosity is directly consistent theoretical expectations that decreased effective population sizes would lower heterozygosity to $H_S=0.00131$.    

At non-synonymous sites, however, there are no closed-form solutions for how $H_N$ would decay under reduced population sizes.  We observe  $H_N=0.000490$ in the Wrangel Island Mammoth, 95\% of $H_N=0.000506$ in the Oimyakon mammoth (Table \ref{HetNS}).  To determine whether such results could be caused by accumulation of nearly-neutral variation, we simulated population trajectories estimated using PSMC.   We were able to qualitatively confirm results that population trajectories from PSMC with previously described mutation rates and selection coefficients can lead to an accumulation of detrimental alleles in populations.  However, the magnitude of the effects is difficult to fit precisely.   The simulations show a mean $H_S=0.00148$ and $H_N=0.000339$ in Oimyakon and $H_S=0.00126$ and $H_N=0.000295$ for the Wrangel Island Mammoth  (Figure \ref{MammothSims}).  In simulations, we estimate  $H_N/H_S=0.229$ both for the Oimyakon mammoth and directly after the bottleneck, but $H_N/H_S=0.233$ in the Wrangel Island Mammoth at the time of the Wrangel Island mammoth.  These numbers are less than empirical observations of $H_N/H_S=0.370$ (Table \ref{HetNS}).  Several possibilities might explain the observed disparity between precise estimates from simulations versus the data.  The simulations may be particularly sensitive to perturbations from PSMC population levels or time intervals.  Similarly, selection coefficients that differ from the gamma distribution previously estimated for humans might lead to greater or lesser changes in small populations.  Additionally, an acceleration in generation time on Wrangel Island is conceivable, especially given the reduced size of Wrangel Island mammoths \citep{Vartanyan1993}.   Finally, positive selection altering nucleotide variation on the island or the mainland could influence diversity levels.

Founder effects during island invasion sometimes alter genetic diversity in populations.  However, it is unlikely that a bottleneck alone could cause an increase in $H_N/H_S$.  There is no evidence in effective population sizes inferred using PSMC to suggest a strong bottleneck during Island colonization \citep{Palkopoulou2015}.  The power of such genetic analyses may be limited, but these results are in agreement with paleontological evidence showing no phenotypic differentiation from the mainland around 12,000 years ago followed by island dwarfism much later \citep{Vartanyan1993}.  During glacial maxima, the island  was fully connected to the mainland, becoming cut off as ice melted and sea levels rose.  The timing of separation between the island and mainland lies between 10,000 years and 14,000 years before present \citep{Vartanyan1993, Elias1996,Lozhkin2001,Vartanyan2008}, but strontium isotope data for mammoth fossils suggests full isolation of island populations was not complete until 10,000-10,500 years ago \citep{arppe2009}.   Forward simulations suggest that hundreds of generations at small $N_e$ are required for detrimental mutations to appear and accumulate in the population.  These results are consistent with recent theory suggesting extended bottlenecks are required to diminish population fitness  \citep{Balick2015}.  Thus, we suggest that a bottleneck alone could not produce the accumulation of $H_N/H_S$ that we observe.  

\Emax {} specimen, Maya shows an independent population decline in the past 100,000 years, with current estimates of $N_e=~1000$ individuals (Figure \ref{BadPSMC}).  This specimen shows a parallel case of declining population sizes in a similar species of elephantid.  Maya houses hemizygous deletions in similar numbers with the Wrangel Island Mammoth.  However, the number of stop codons and homozygous deletions is intermediate in comparison with the Oimyakon and Wrangel mammoths (Table \ref{MutTypes}).  It is possible that Indian elephants, with their recently reduced population sizes may be subject to similar accumulation of detrimental mutations, a prospect that would need to be more fully addressed in the future using population genomic samples for multiple individuals or timepoints and more thorough analyses.  

\section*{Discussion}
\subsection*{Nearly neutral theories of genome evolution}
Nearly-neutral theories of genome evolution have attempted to explain the accumulation of genome architecture changes across taxa \citep{LynchBook}.  Under such models, mutations with selection coefficients less than the nearly neutral threshold will accumulate in genomes over time. Here, we test this hypothesis using data from a woolly mammoth sample from just prior to extinction.   We observe an excess of retrogenes, deletions, amino acid substitutions, and premature stop codons in woolly mammoths on Wrangel Island.    Given the long period of isolation and extreme population sizes observed in pre-extinction mammoths on Wrangel Island, it is expected that genomes would deteriorate over time.  These results offer genetic support for the nearly-neutral theory of genome evolution, that under small effective population sizes, detrimental mutations can accumulate in genomes.  Independent analysis supporting a reduction in nucleotide diversity across multiple individuals at MHC loci suggests a loss of balancing selection further support the hypothesis that detrimental variants accumulated in small populations  \citep{pevcnerova2016}.  

We observe two independent loss-of-function mutations in the Wrangel Island mammoth at the locus of \Fox.  One mutation removes the entire gene sequence via a deletion, while the other produces a frameshift in the CDS.  Based on phenotypes observed in mouse models, these two independent mutations would result in a satin fur coat, as well as gastric irritation \citep{Verzi2008}.  Many phenotypic screens search for homozygous mutations as causative genetic variants that could produce disease.  More recently, it has been proposed that the  causative genetic variation for disease phenotypes may be heterozygous non-complementing detrimental mutations \citep{Thornton2013}.  These data offer one case study of independent non-functionalizing mutations in a single individual, genetic support for independent non-functionalizing mutations at a single locus.  Woolly mammoth outer hairs house multiple medullae, creating a stiff outer coat that may have protected animals from cold climates \citep{Tridico2014} (though see \cite{chernova2015} for alternative interpretations).  Putative loss of these medullae through loss of \Fox {} could compromise this adaptation, leading to lower fitness.     

\subsection*{Island specific changes}
One of the two specimens comes from Wrangel Island, off the northern coast of Siberia.   This mammoth population  had been separated from the mainland population for at least 6000 years after all mainland mammoths had died off.  Prior to extinction, some level of geographic differentiation combined with differing selective pressures led to phenotypic differentiation on Wrangel island \citep{Vartanyan1993}.  Island mammoths had diminished size, but not until 12,000 years ago when mainland  populations had reduced and ice sheets melted \citep{Vartanyan1993}.  One possible explanation for the poor fit of simulations is that generation time may have decreased.  Previous work suggested a very high mutation rate for woolly mammoths based on comparisons between island and mainland mammoths.  It is possible that an acceleration in generation times could cause the accumulation of more mutations over time, and that the real mutation rate is similar to humans ($1-2\times10^{-8}$ \citep{Scally2012} rather than $3.8\times10^{-8}$ \citep{Palkopoulou2015}).  Such changes would be consistent with island dwarfism being correlated with shorter generation times, and would explain the unusually high mutation rate estimate for mammoths based on branch shortening observed in \citep{Palkopoulou2015}.   

We observe large numbers of pseudogenized olfactory receptors in the Island mammoth.  Olfactory receptors evolve rapidly in many mammals, with high rates of gain and loss \citep{Nei2008}.  The Wrangel island mammoth has massive excess even compared to the mainland mammoth.  Wrangel island had different flora compared to the mainland, with peat and sedges rather than grasslands that characterized the mainland \citep{Lozhkin2001}.  The island also lacked large predators present on the mainland.   It is possible that island habitats created new selective pressures that resulted in selection against some olfactory receptors.     Such evolutionary change would echo gain and loss of olfactory receptors in island \emph{Drosophila} \citep{stensmyr2003}.  In parallel, we observe a large number of deletions in major urinary proteins in the island mammoth.  In Indian elephants \Emax, urinary proteins and pheromones ellicit behavioral responses including mate choice and social status \citep{rasmussen2003}.  It is possible that coevolution between urinary proteins, olfactory receptors, and vomeronasal receptors led to a feedback loop, allowing for rapid loss in these related genes.  It is equally possible that urinary peptides and olfactory receptors are not essential and as such they are more likely to fall within the nearly neutral range \citep{Nei2008}. Either of these hypotheses could explain the current data. 

\subsection*{Implications for conservation genetics}
Many factors contributed to the demise of woolly mammoths in prehistoric times.  Climate change led to receding grasslands as forests grew in Beringia and North America and  human predation placed a strain on already struggling populations \citep{Nogues2008}.    Unlike many cases of island invasion, Wrangel Island mammoths would not have continuous migration to replenish variation after mainland populations went extinct.  Under such circumstances, detrimental variation would quickly accumulate on the  island.   The putatively detrimental variation observed in these island mammoths, with the excess of deletions, especially recessive lethals may also have limited survival of these struggling pre-extinction populations.  Climate change created major limitations for mammoths on other islands \citep{Graham2016}, and these mammoths may have struggled to overcome similar selective pressures.

Many modern day species, including elephants, are threatened or endangered.  Asiatic cheetahs are estimated to have fewer than 100 individuals in the wild \citep{Hunter2007}.   Pandas are estimated to have 1600 individuals living in highly fragmented territories \citep{Wang2009}. Mountain Gorilla population census sizes have been estimated as roughly 300 individuals, similar to effective population sizes for pre-extinction mammoths  \citep{guschanski2009}.   If nearly neutral dynamics of genome evolution affect contemporary endangered species, detrimental variation would be expected in these genomes.  With single nucleotide changes, recovered populations can purge detrimental variation in hundreds to thousands of generations, returning to normal genetic loads \citep{Balick2015}.  However, with deletions that become fixed in populations, it is difficult to see how genomes could recover quickly.  The realm of back mutations to reproduce deleted gene sequences will be limited or impossible.  Although compensatory mutations might conceivably correct for some detrimental mutations, with small effective population sizes, adaptation through both new mutation and standing variation may be severely limited \citep{Pennings2006}.   Thus we might expect genomes affected by genomic meltdown to show lasting repercussions that will impede population recovery.    

\section*{Methods}
\subsection*{Genome Sequence Alignments}
We used previously aligned bam files from ERR852028 (Oimyakon, 11X) and  ERR855944 (Wrangel, 17X) (Table \ref{SRANumbers}) \citep{Palkopoulou2015} aligned against the \Lafr {} 4.0 reference genome  (available on request from the Broad Institute - \href{mailto:vertebrategenomes@broadinstitute.org}{vertebrategenomes@broadinstitute.org}; \url{https://www.broadinstitute.org/scientific-community/science/projects/mammals-models/elephant/elephant-genome-project}).  We also aligned 33X coverage of sequencing reads for one modern \Emax {} genome Maya (previously described as ``Uno") using bwa 0.7.12-r1044 \citep{bwa}, with parameters set according to \cite{Palkopoulou2015}  \texttt{bwa aln -l 16500 -o 2 -n 0.01}.  The \Emax {} sample, previously labeled in the SRA as ``Uno", is from Maya, a former resident of the San Diego Zoo wild-born in Assam, India, North American Studbook Number 223, Local ID \#141002 (O. Ryder, personal communication).  We were not able to use two other mammoth sequences are publicly available, M4 and M25 from Lynch et al. \citep{Lynch2015}.  These sequences display abnormal PSMC results (Figure \ref{BadPSMC}), high heterozygosity (Figure \ref{HetSites}), and many SNPs with asymmetrical read support (Figure \ref{AsymFig}).  The unrealistically high heterozygosity as well as abnormal heterozygote calls raise concerns with respect to sequence quality. For further description, please see Supporting Information. 

\subsection*{Synonymous and nonsynonymous substitutions}
We used the GATK pipleine \citep{GATK}  v3.4-0-g7e26428 to identify SNPs in the aligned sequence files for the Oimyakon and Wrangel Island mammoths. We identified and realigned all indel spanning reads according to the standard GATK pipeline.  We then identified all SNPs using the Unified Genotyper, with output mode set to emit all sites.  We used all CDS annotations from cDNA annotations from \Lafr {} r3.7 with liftover coordinates provided for \Lafr {} 4.0 to identify SNPs within coding sequences.  We identified all stop codons, synonymous substitutions, and non-synonymous substitutions for the Wrangel Island and Oimyakon mammoths at heterozygous and homozygous sites.

\subsection*{Retrogenes}
We aligned all reads from the mammoth genome sequencing projects ERR852028 (Oimyakon) and  ERR855944 (Wrangel) (Table \ref{SRANumbers}) against elephant cDNA annotations from \Lafr {} r3.7.  Sequences were aligned using bwa 0.7.12-r1044 \citep{bwa}, with parameters set according to \citep{Palkopoulou2015}  \texttt{bwa aln -l 16500 -o 2 -n 0.01} in order to account for alignments of damaged ancient DNA.  We then collected all reads that map to exon-exon boundaries with at least 10 bp of overhang.   Reads were then filtered against aligned genomic bam files produced by Palkopoulou et al \citep{Palkopoulou2015}, discarding all exon-exon junction reads that have an alignment with equal or better alignments in the genomic DNA file.  We then retained all putative retrogenes that showed signs of loss for two or more introns, using only cases with 3 or more exon-exon junction reads.  

\subsection*{Deletions}
We calculated coverage depth using samtools \citep{samtools} with a quality cutoff of -q 20.   We then implemented change point analysis  \citep{Yao1988} in 20 kb windows.  Change point methods have been commonly used to analyze microarray data and single read data for CNVs \citep{olshen2004, chiang2009, Niu2012}  The method seeks compares the difference in the log of sum of the squares of the residuals with one regression line vs. two regression lines \citep{Yao1988}.  The test statistic follows a chi-squared distribution with a number of degrees of freedom determined by the number of change-points in the data, in this case $df=1$.  We required significance at a Bonferroni corrected  p-value of 0.05 or less.   We allowed for a maximum of one CNV tract per window, with minimum of 1 kb and maximum of 10 kb (half the window size) with a 100 bp step size.  We did not attempt to identify deletions smaller than 1 kb due to general concerns of ancient DNA sequence quality, limitations to assess small deletions in the face of stochastic coverage variation, and concerns that genotype calls for smaller deletions might not be as robust to differences in coverage between the two mammoths.  Sequences with 'N's in the reference genome did not contribute to change point detection.  We excluded all deletions that were identified as homozygous mutations in both mammoths and in \Emax {} specimen Maya, as these suggest insertion in the \Lafr {} reference rather than deletion in other elephantids.  To determine the effects that coverage differences would have on deletions, we downsampled the sequence file for the Wrangel Island mammoth using samtools to 11X coverage, using chromosome 1 as a test set.  We observe a reduction in the number of deletions for chromosome 1 from 1035 deletions to 999 deletions, resulting in an estimated false negative rate of 0.5\% at reduced coverage for deletions greater than 1 kb.  Highly diverged haplotypes with greater than 2\% divergence might prevent read mapping and mimic effects of deletions, but this would require divergence times  within a species that are greater than the divergence between mammoths and \Lafr.  Mutations were considered homozygous if mean coverage for the region was less than 10\% of the background coverage level.  Otherwise it was considered to be heterozygous.  These methods are high-throughput, and it is possible that multiple small homozygous  deletions interspersed with full coverage sequences might mimic heterozygote calls.  Whether such mutations might meet the conditions for significant change-point detection would depend on the deletion length, placement, and background coverage level.  

\subsection*{Demography}
We identified SNPs that differentiate Mammoth genomes from the reference {} using samtools mpileup (options -C50 -q30 -Q30), and bcftools 1.2 consenus caller (bcftools call -c).  The resulting vcf was converted to fastq file using bcftools vcf2fq.pl with a mimimum depth of 3 reads and a maximum depth of twice the mean coverage for each genome.  Sequences were then converted to psmc fasta format using fq2psmcfa provided by psmc  0.6.5-r67.  We then ran psmc with 25 iterations (-N25), an initial ratio of $\theta$/$\rho$ of 5 (-r5), and parameters 64 atomic time intervals and 28 free parameters (\texttt{-p "4+25*2+4+6"}) as was done in previous analysis of woolly mammoths \citep{Palkopoulou2015}.    Effective population sizes and coalescence times were rescaled using previously estimated mutation rates of $3.8\times10^{-8}$.  Using the population size estimates from PSMC, we calculated the expected reduction in heterozygosity at synonymous sites according to $(1-\frac{1}{2N})^t$ for each time period in PSMC output.  We compared the number of deletions, number of premature stop codons, proportion affecting gene sequences, and number of putative retrogenes between the two mammoth genomes using chi squared tests. 
 
 \subsection*{Simulations}
To determine expectations of sequence evolution at non-synonymous sites under population crash, we ran simulations using SLiM v. 2.0 population genetic software \citep{Messer2013}.  We modeled two classes of sites: neutral and detrimental.  For detrimental mutations we used a gamma distributed DFE with a mean of -0.043 and a shape parameter of 0.23 as estimated for humans \citep{EyreWalker2006}, assuming a dominance coefficient of 0.5 and free recombination across sites.  Mutation rates were set as $3.8\times10^{-8}$ based on previously published estimates \citep{Palkopoulou2015}.  The trajectory of population sizes was simulated according to estimates from PSMC, omitting the initial and final time points from PSMC, which are often subject to runaway behavior.   We then simulated the accumulation of $H_N/H_S$ in the Wrangel Island Mammoths. Simulations were run with a burn-in of 100,000 generations.  We simulated 460 replicates of haplotypes with 100 sites for each mutation class.

\subsection*{Gene Ontology}
To gather a portrait of functional categories captured by deletions, retrogenes, and stop codons, we identified all mouse orthologs based on ENSEMBL annotations for \Lafr {} 3.7 for affected gene sequences.  We then used DAVID gene ontology analysis with the clustering threshold set to `Low' (http://david.ncifcrf.gov/; Accessed April 2016) \citep{DAVID1,DAVID2}.  Tables \ref{GOStops}-\ref{GORetrosWrangel} include all functions overrepresented at an EASE enrichment cutoff of 2.0.  Full gene ontology data is included in Supplementary Information.

\section*{Acknowledgements}
The authors would like to thank Oliver Ryder and Lynn Tomsho for sharing information about \Emax {} specimen Maya (also known as ``Uno") and Charles Marshall for helpful discussions about woolly mammoths.  The authors would also thank Jeremy Johnson at the Broad Institute, who kindly provided \LafrFull {} r.4 genome assembly and liftover files.  We thank Vincent Lynch and Webb Miller for helpful discussions of the data presented in Lynch et al. 2015. We thank three anonymous reviewers for their comments that improved the manuscript. RRLR and MS are funded by grant R01-GM40282 from the National Institutes of Health
to Montgomery Slatkin. The funders had no role in study design, data collection and analysis, decision to publish, or preparation of the manuscript.


\clearpage
\bibliographystyle{MBE}
\bibliography{MammothRevisions}
\clearpage

\section*{Supporting Data Files}
Text S1 - Supporting Text \\
Table S1 - Non synonymous and synonymous changes in Wrangel and Oimyakon mammoths \\
Table S2 - Gene ontology categories for premature stop codons in mammoths \\
Table S3 - Deletions identified in mammoth genomes \\ 
Table S4 - Shared deletions in mammoth genomes \\
Table S5 - Gene ontology for deleted exons \\
Table S6 - Gene ontology for retrogenes in the Oimyakon mammoth \\ 
Table S7 - Gene ontology for retrogenes in the Wrangel Island mammoth \\
Table S8 - SRA and ENA Identifiers for Mammoth and Elephant Sequence Data \\
Table S9 - Heterozygous sites per 10 kb \\
Table S10 - Asymmetrical Support \\
Figure S1 - Coverage depth at the RFK locus \\
Figure S2 - Coverage depth at the FOXQ1 locus  \\ 
Figure S3 -  Simulations for heterozygosity at synonymous and non-synonymous sites  \\
Figure S4 - PSMC results \\ 
Figure S5 - Heterozygosity for mammoth and elephant samples. \\
Figure S6 - Asymmetric SNPs \\
Figure S7 - Asymmetric SNPs, excluding damage \\
SuppFiles.zip  - Data release archive \\

\clearpage
\begin{center}
\begin{threeparttable}
\caption{\label{MutTypes} Mutations Identified in Mammoth Genomes  }
\begin{tabular}{lrrr}
 Mutation & Oimyakon & Wrangel & Maya \\
 \hline
 Deletions  & 21346$^1$ &  27228 & 28095 \\
 Retrogenes &	2130 & 2853 &  1575  \\
Genes with exons deleted &   1115$^1$ & 1628 & 3427   \\
Stop Codons & 503$^2$ &  819  &  450 \\ 
Stop Codons, excluding damage & 377 & 645 & 401 \\ 
\hline
\end{tabular}
\begin{tablenotes}
\item[1] Corrected for false negative rate of 0.5\% in heterozygotes 
\item[2] Corrected for false negative rate of 30\% at heterozygous sites established by Palkopoulou et al 2015.
\end{tablenotes}
\end{threeparttable}
\end{center}

 \clearpage
 \begin{center}
\begin{threeparttable}
\caption{\label{HetNS}Non-synonymous and Synonymous Heterozygosity}
\center
\begin{tabular}{lll}
 & Wrangel & Oimyakon$^1$ \\
 \hline
$H_S\pm2\sigma$ & $0.00130\pm0.00002$  & $0.00161\pm0.00002$ \\
$H_N\pm2\sigma$ & $0.000490\pm0.000012$& $0.000506\pm0.000012$ \\
$H_N$/$H_S$ & 0.370 & 0.314 \\
 \hline
\end{tabular}
\begin{tablenotes} \item[1] Oimyakon corrected for false negative rate of 30\% established by Palkopoulou et al 2015.\\		   
\end{tablenotes}
\end{threeparttable}
\end{center}

\clearpage

\clearpage

\clearpage
\section*{Supporting Information}
\renewcommand{\thefigure}{S\arabic{figure}}
\renewcommand{\thetable}{S\arabic{table}}
\setcounter{figure}{0}
\setcounter{table}{0}
\setcounter{page}{1}


\subsection*{Analysis of samples M4 and M25}
We aligned all major runs in the SRA for two \Mprim {} specimens previously published, M4 and M25 (Table \ref{SRANumbers}) \citep{Lynch2015}.  As a comparison for sequence quality, we also aligned and analyzed reads for one female \Emax {} specimen, Maya, sequenced and processed in the same study.   Previously published sequences for all three elephantids were aligned to the \Lafr {} r.4.0 reference genome using bwa 0.7.12-r1044  \citep{bwa}, with parameters set according to \citep{Palkopoulou2015}  \texttt{bwa aln -l 16500 -o 2 -n 0.01}.  Indels were identified and realigned using GATK as defined above.  We then generated all SNPs using samtools mpileup (-C50 -u -g) and consensus fastq was generated using bcftools consensus caller (bcftools call -c) and bcftools vcf2fq.pl with a minimum depth threshold of 3 reads and a maximum depth of twice the mean coverage for each genome.   Resulting fastq files were converted to psmcfa using the PSMC toolkit  \citep{PSMC}.   We then ran PSMC  \citep{PSMC} exactly as described in \cite{Palkopoulou2015}, with 64 time intervals, \texttt{(-p "4+25*2+4+6")}.  

Demographic inference for mammoth samples from Oimyakon and Wrangel Island \citep{Palkopoulou2015} show $N_e \leq 25,000$ (Figure \ref{BadPSMC}).  Analysis of samples M25 and M4 suggests $N_e$ in the range of $10^{10}$-$10^{11}$ over the history of woolly mammoths (Figure \ref{BadPSMC}), a result that is inconsistent with estimates based on mtDNA \citep{Barnes2007} or habitat availability \citep{Nogues2008}.   Demographic inference for Maya the elephant yields $N_e< 20,000$, with a  bottleneck event roughly 200,000 years ago.

Given the inconsistencies in the M4 and M25 results, we examined heterozygosity data more directly for each of the samples, using chromosome 1 as an example dataset.  We calculated heterozygosity for 10 kb windows in each mammoth and elephant sample.  M4 and M25 both display high heterozygosity.  We observe 30 heterozygous sites per 10 kb window in M4, and 38 heterozygous sites per 10 kb window in M25.   These numbers are 2-3 fold higher than the observed mean of 11-14 sites per 10 kb window in Wrangel, Oimyakon, and Maya (Table \ref{HetNumbers}; Figure \ref{HetSites}).  The abnormally high heterozygosity is likely to explain abnormal estimates of $N_e$ from PSMC. We then examined support for heterozygous SNP calls, using the first 5000 SNPs on chromosome 1 as a test set.  If sites are truly heterozygous, there should be symmetrical support for each base by site.  We identified sites with significantly skewed support in a binomial test.  Mammoth specimens M4 and M25 from \citep{Lynch2015} have an excess of SNPs with significantly asymmetrical support compared to the Oimyakon and Wrangel mammoths, as well as Maya the elephant (Table \ref{AsymNumbers}; Figure \ref{AsymM4}-\ref{AsymMaya}).  There is a greater number of asymmetric sites that favor the reference allele than the non-reference allele in both M4 and M25 (Table \ref{AsymNumbers}; Figure \ref{AsymM4}-\ref{AsymM25}).  Such asymmetry would be expected if some other elephantid DNA had contaminated these two samples, or if in vitro recombination occurred between barcodes during PCR amplification or sequencing.  Removing A/G and T/C mutations did not correct the pattern, suggesting that these results are not a product of differences in damage for archaic samples (Figure \ref{AsymDamageFig}).  Multiple mammoths were sequenced in the lab, only some of which have been published (http://mammoth.psu.edu/moreThanOne.html; accessed June 18, 2016).  We are currently unable to examine all potential sources of contamination.  These results left us concerned for the quality of the sequences. Hence, we did not include the two mammoth specimens M4 and M25 in the current genomic analysis of deletions, retrogenes, stop codons, or amino acid substitutions.


\clearpage
\begin{center}
\begin{threeparttable}
\caption{\label{ChiSqTable}Non-synonymous and synonymous sites}
\center
\begin{tabular}{llrr}
 & & Wrangel & Oimyakon$^1$ \\
 \hline
Heterozygous & Non-synonymous & 12784 & 9445 \\
 & Synonymous &  10231 &  8913 \\
Homozygous & Non-synonymous & 16149 & 13447 \\
 & Synonymous & 21842 & 18950  \\
 \hline
\end{tabular}
\begin{tablenotes} \item[1] Raw numbers, without correction for changes in coverage. \\		   
\end{tablenotes}
\end{threeparttable}
\end{center}

\clearpage

\begin{table}
\center
 \caption{DAVID Gene ontology for premature stop codons in the Wrangel Island Mammoth}
\begin{tabular}{llr}
Specimen&Function & EASE score\\
     \hline
Oimyakon & Olfactory receptors & 4.1\\
Wrangel & Olfactory receptors & 9.1 \\
& Ankyrin domains & 1.6  \\
\hline
\end{tabular}
\label{GOStops}
\end{table}

\clearpage
\begin{center}
\begin{threeparttable}
\caption{\label{HetDels} Heterozygosity for Deletions Identified in Elephantid Genomes  }
\begin{tabular}{llrrr}
  & & Oimyakon & Wrangel & Maya\\
 \hline
All & Homozygous$^1$  & 5035 & 6147 & 5141   \\
&Hemizygous & 16223 & 21081 & 22954 \\
Exon Deletions & Homozygous & 136 & 173 & 165 \\
& Hemizygous & 1347 & 1985 & 4248 \\
 \hline
\end{tabular}
\begin{tablenotes}
\item[1] $\leq 10\%$ coverage  
\end{tablenotes}
\end{threeparttable}		
		
\end{center}

\clearpage
\begin{center}
\begin{threeparttable}
\caption{\label{Shared} Shared Deletions Identified in Mammoth Genomes  }
\begin{tabular}{lr}
  & Mutations \\
 \hline
Homozygous$^1$  & 3001   \\
 Heterozygous  & 9581 \\
 Mixed & 877 \\
 Total & 13459 \\
 \hline
\end{tabular}
\begin{tablenotes}
\item[1] $\leq 10\%$ coverage  
\end{tablenotes}
\end{threeparttable}		
		
\end{center}

 \clearpage

\begin{table}
\center
 \caption{DAVID Gene ontology for deleted exons}
\begin{tabular}{llr}
Specimen & Function & EASE score\\
     \hline
Oimyakon & Cell junction & 4.6 \\
& Neurons & 3.42 \\
& Zinc fingers & 3.41 \\
& Aldo/keto metabolism & 3.10 \\
& Calcium dependent transport & 2.91\\
& DNA damage & 2.85 \\
& Transcription regulation & 2.71 \\
& Development & 2.66 \\
Wrangel & Major Urinary proteins & 7.95  \\
 & Pleckstrins & 5.49 \\
 & Transcription regulation & 4.86 \\
 & Cellular transport & 3.51 \\
 & DNA damage & 3.34 \\
 & Chromatin regulation & 3.15 \\
 & \emph{Hox} domains & 3.06 \\
 & Development & 2.75 \\
\hline
\end{tabular}
\label{GODels}
\end{table}

\clearpage

\begin{table}s
\center
 \caption{DAVID Gene ontology for retrogenes in the Oimyakon Mammoth}
\begin{tabular}{lr}
Function & EASE score\\
     \hline
Ribosome & 6.3 \\
Post translational modification & 4.4 \\
Lipoproteins & 3.4 \\
Spliceosome & 3.1 \\
RNA binding & 2.6 \\
Lipoprotein metabolism & 2.2 \\
Nucleolus & 2.0 \\
Glutamine metabolism & 1.9 \\
Aspartate metabolism & 1.8 \\
Starch and drug metabolism & 1.7 \\
Proteasome & 1.6 \\
Translation initiation & 1.6 \\
\hline
\end{tabular}
\label{GORetrosOimyakon}
\end{table}

\clearpage

\begin{table}
\center
 \caption{DAVID Gene ontology for retrogenes in the Wrangel Island Mammoth}
\begin{tabular}{lr}
Function & EASE score\\
     \hline
Ribosome  & 8.3 \\
Ubl conjugation & 6.8 \\
Spliceosome & 4.3 \\
Translation initiation & 2.8 \\
Lipoprotein & 2.6 \\
Nuclear body & 2.3 \\
Cytoskeleton & 2.0 \\
Aminoacylation & 1.8 \\
HEAT elongation & 1.6 \\
RNA splicing & 1.6 \\
\hline
\end{tabular}
\label{GORetrosWrangel}
\end{table}

\clearpage

\begin{table}
\center
 \caption{SRA and ENA Identifiers for Mammoth and Elephant Sequence Data  }
\begin{tabular}{rl}

 Specimen & Database ID\\
     \hline
Oimyakon   &  ERR852028 \\  
Wrangel & ERR855944 \\
Maya & SRX1015606 \\ 
& SRX1015608 \\
M4 & SRX1015711 \\ 
& SRX1015712 \\
& SRX1015714 \\
& SRX1015715 \\
& SRX1015717 \\
& SRX1015679 \\
& SRX1015671 \\
& SRX1015640 \\
& SRX1015634 \\
& SRX1015625 \\
M25 & SRX1015733 \\
& SRX1015732 \\ 
& SRX1015729 \\
& SRX1015727 \\
& SRX1015726 \\ 
\hline
\end{tabular}
\label{SRANumbers}
\end{table}

\clearpage

\begin{table}
\center
 \caption{Heterozygous sites per 10 kb  }
\begin{tabular}{rl}
    Hets & Specimen\\
     \hline
 12   &  Wrangel\\
 14  &  Oimyakon  \\
 11 & Maya \\
 30 &  M4  \\
 38 &  M25  \\  
\hline
\end{tabular}
\label{HetNumbers}
\end{table}

\clearpage

\begin{table}
\center
 \caption{Asymmetrical Support }
\begin{tabular}{rrrl}

    Asymm SNPs & Favor Ref & Favor Alt & Specimen\\
     \hline
   498 & 166 & 332 &  Wrangel\\
 217 & 59 & 158 &  Oimyakon  \\
377 & 240 & 137 &  Maya \\
\hline
1355 & 1179 & 176 & M4  \\
 2383 & 1859 & 524  &   M25  \\  
\hline
\end{tabular}
\label{AsymNumbers}
\end{table}

\clearpage
 \begin{figure} 

 \center

\begin{subfigure}[b]{0.35\textwidth}
\includegraphics[scale=0.35]{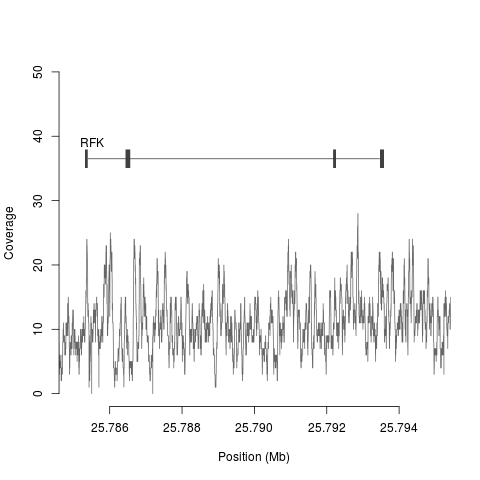}
 \caption{  Oimyakon }
       
    \end{subfigure}
 \begin{subfigure}[b]{0.35\textwidth}
\includegraphics[scale=0.35]{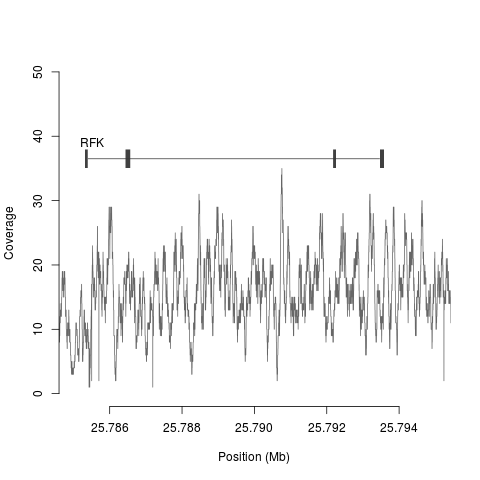}
 \caption{ Wrangel Island } %
       
    \end{subfigure}

\caption{Coverage depth at the \emph{RFK} locus in the A) Oimyakon mammoth and B) Wrangel Island Mammoth.  There is a 50\% reduction in coverage at the first exon of \emph{RFK} in the Wrangel Island mammoth but not in the Oimyakon mammoth. \label{RFKDepth} }
\end{figure}

\clearpage
 \begin{figure} 

 \center

\begin{subfigure}[b]{0.35\textwidth}
\includegraphics[scale=0.35]{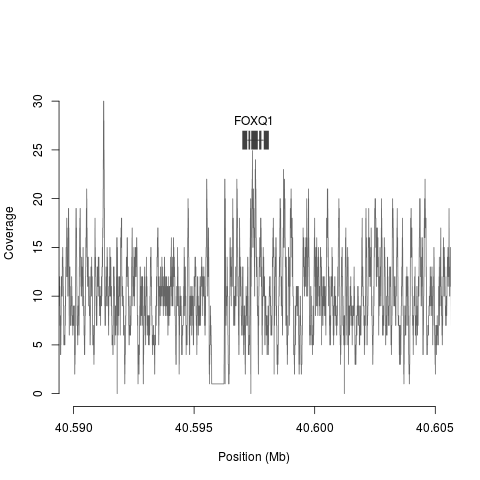}
 \caption{  Oimyakon }
       
    \end{subfigure}
 \begin{subfigure}[b]{0.35\textwidth}
\includegraphics[scale=0.35]{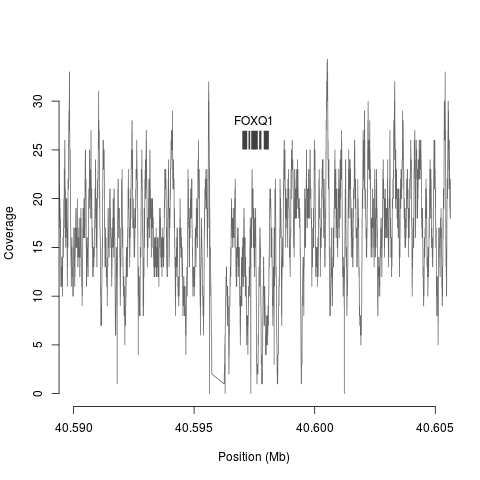}
 \caption{ Wrangel Island } %
       
    \end{subfigure}

\caption{Coverage depth at the \Fox {} locus in the A) Oimyakon mammoth and B) Wrangel Island Mammoth.  There is a 50\% reduction in coverage at \Fox {} in the Wrangel Island mammoth but not in the Oimyakon mammoth.  \label{SatinDepth} }
\end{figure}

\clearpage
 \begin{figure}
 \center
\includegraphics[scale=0.8]{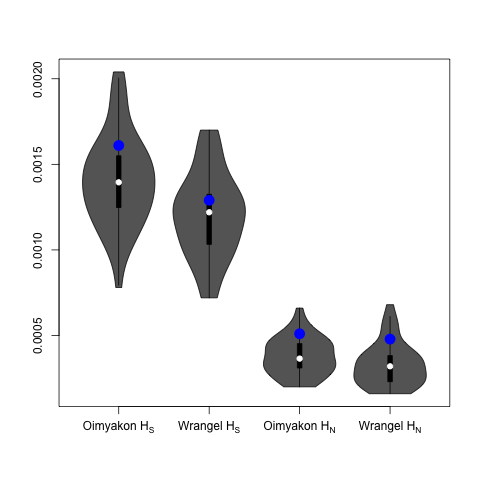}
\caption{ \label{MammothSims} Simulations for heterozygosity at synonymous and non-synonymous sites for the Oimyakon and Wrangel Island mammoths. Black bars show upper and lower quartiles. The white dot is the median.  Grey fields show the full distribution of datapoints. Empirical values for the genome wide average are shown in blue. }
\end{figure}

\clearpage
 \begin{figure}
 \center
\includegraphics[scale=0.8]{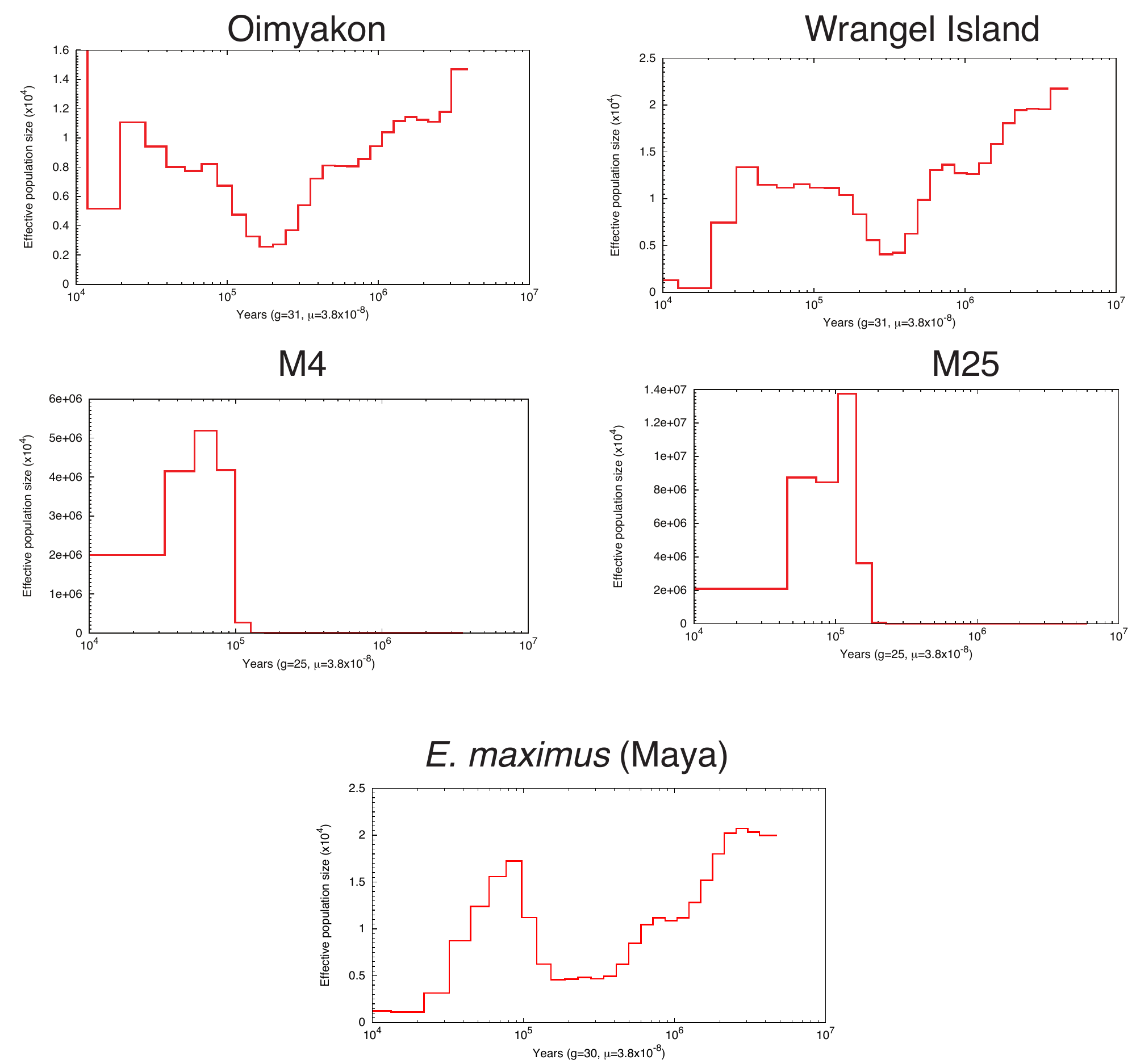}
\caption{ \label{BadPSMC} PSMC results for four woolly mammoths and one elephant.  M4 and M25 both display effective population sizes of $10^{10}$ or higher. }
\end{figure}

\clearpage
 \begin{figure}
 \center
\includegraphics[scale=0.5]{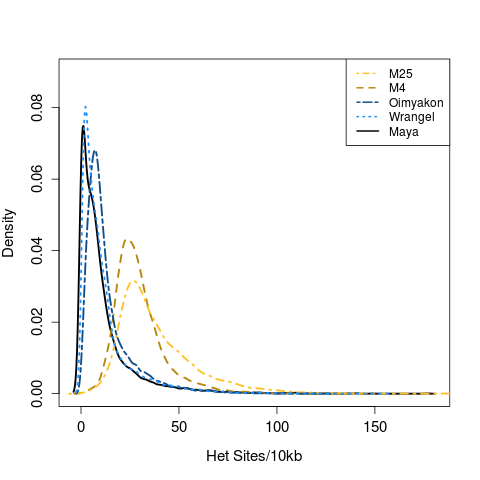}

\caption{\label{HetSites}  Heterozygosity for mammoth and elephant samples. }
\end{figure}

\clearpage
 \begin{figure} 

 \center

\begin{subfigure}[b]{0.35\textwidth}
\includegraphics[scale=0.35]{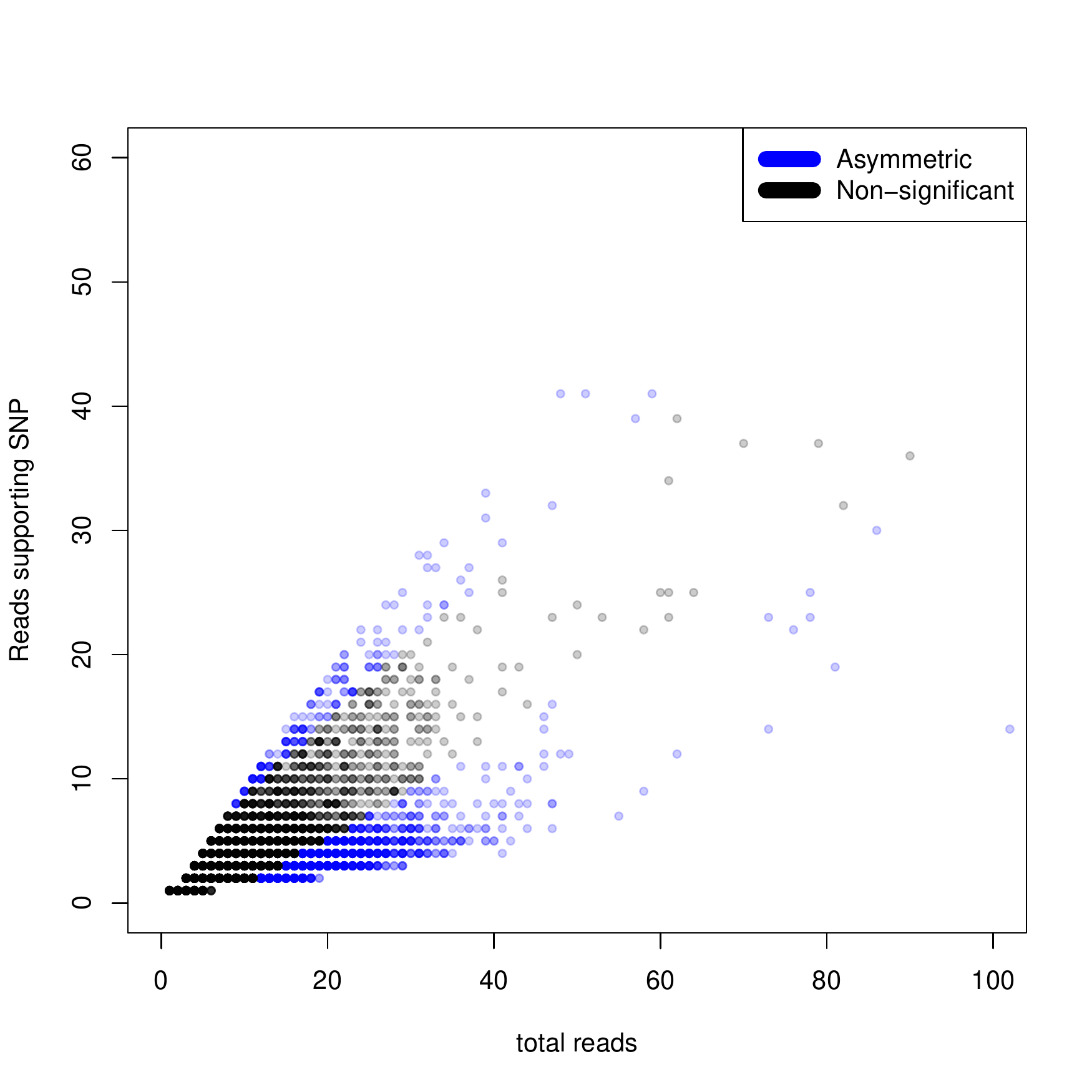}
 \caption{ M4 \label{AsymM4}}
       
    \end{subfigure}
 \begin{subfigure}[b]{0.35\textwidth}
\includegraphics[scale=0.35]{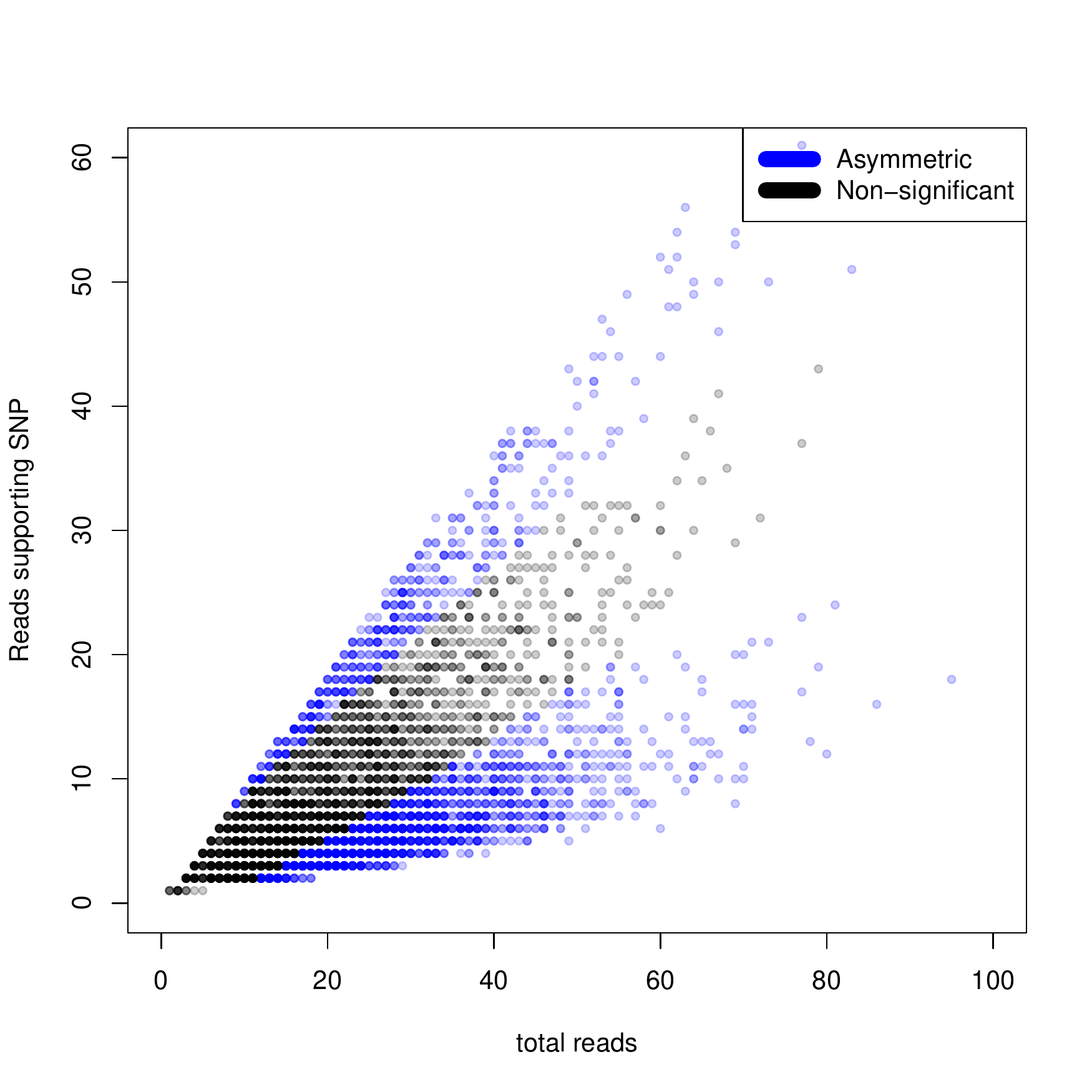}
 \caption{M25 \label{AsymM25}} %
       
    \end{subfigure}

\begin{subfigure}[b]{0.35\textwidth}
\includegraphics[scale=0.35]{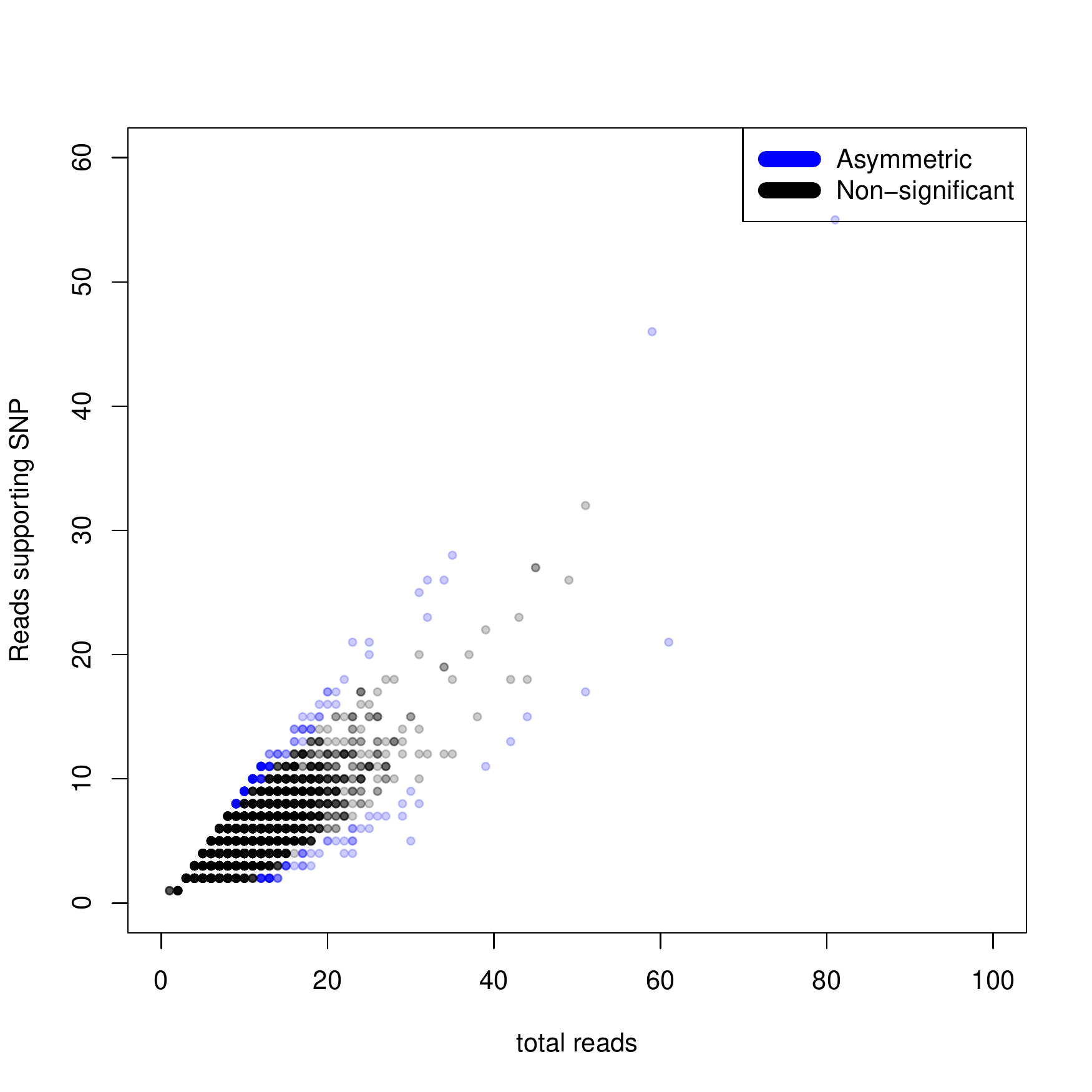}
 \caption{ Oimyakon \label{AsymOimyakon}}
       
    \end{subfigure}
\begin{subfigure}[b]{0.35\textwidth}
\includegraphics[scale=0.35]{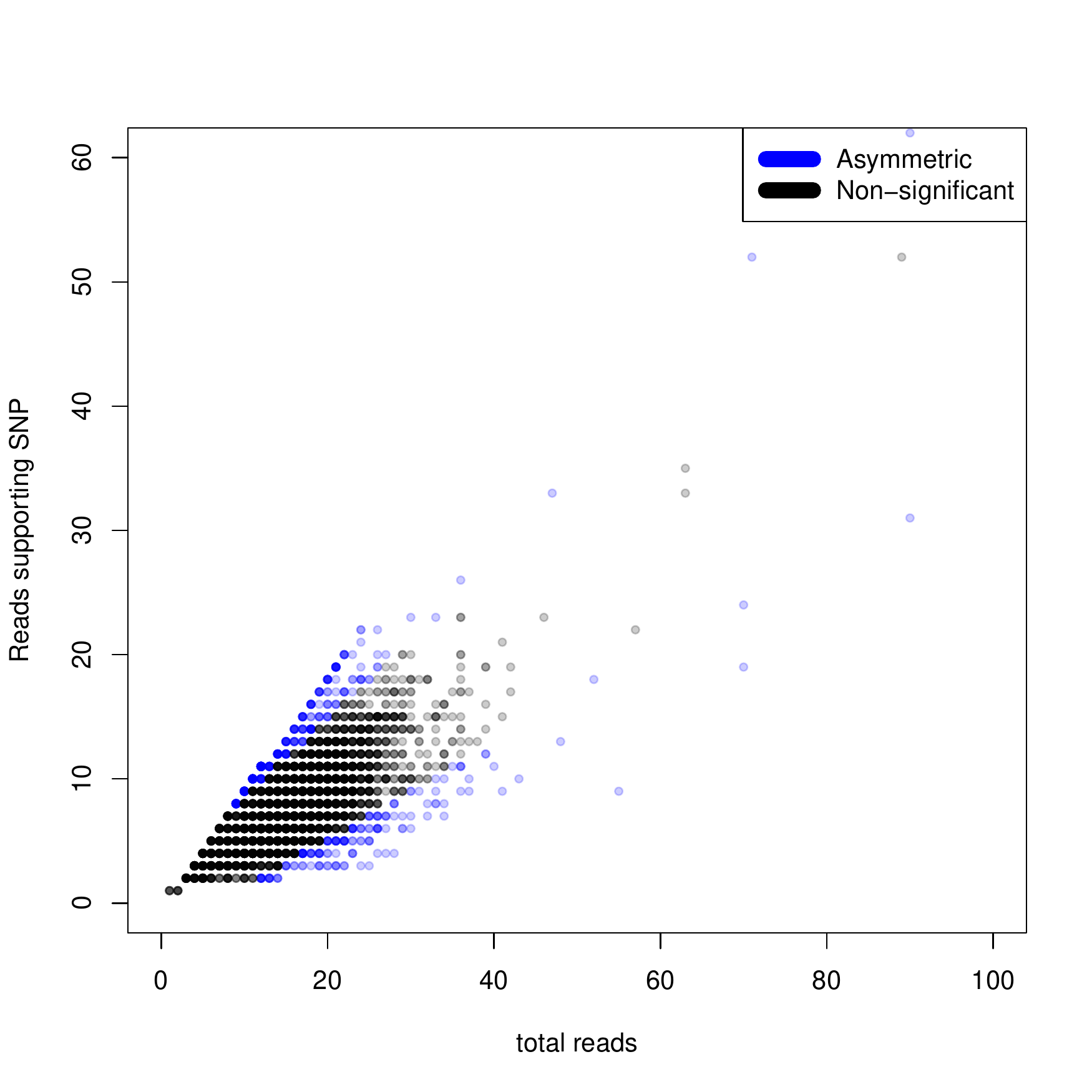}
 \caption{Wrangel  \label{AsymWrangel} Wrangel}
       
    \end{subfigure}

\begin{subfigure}[b]{0.35\textwidth}
\includegraphics[scale=0.35]{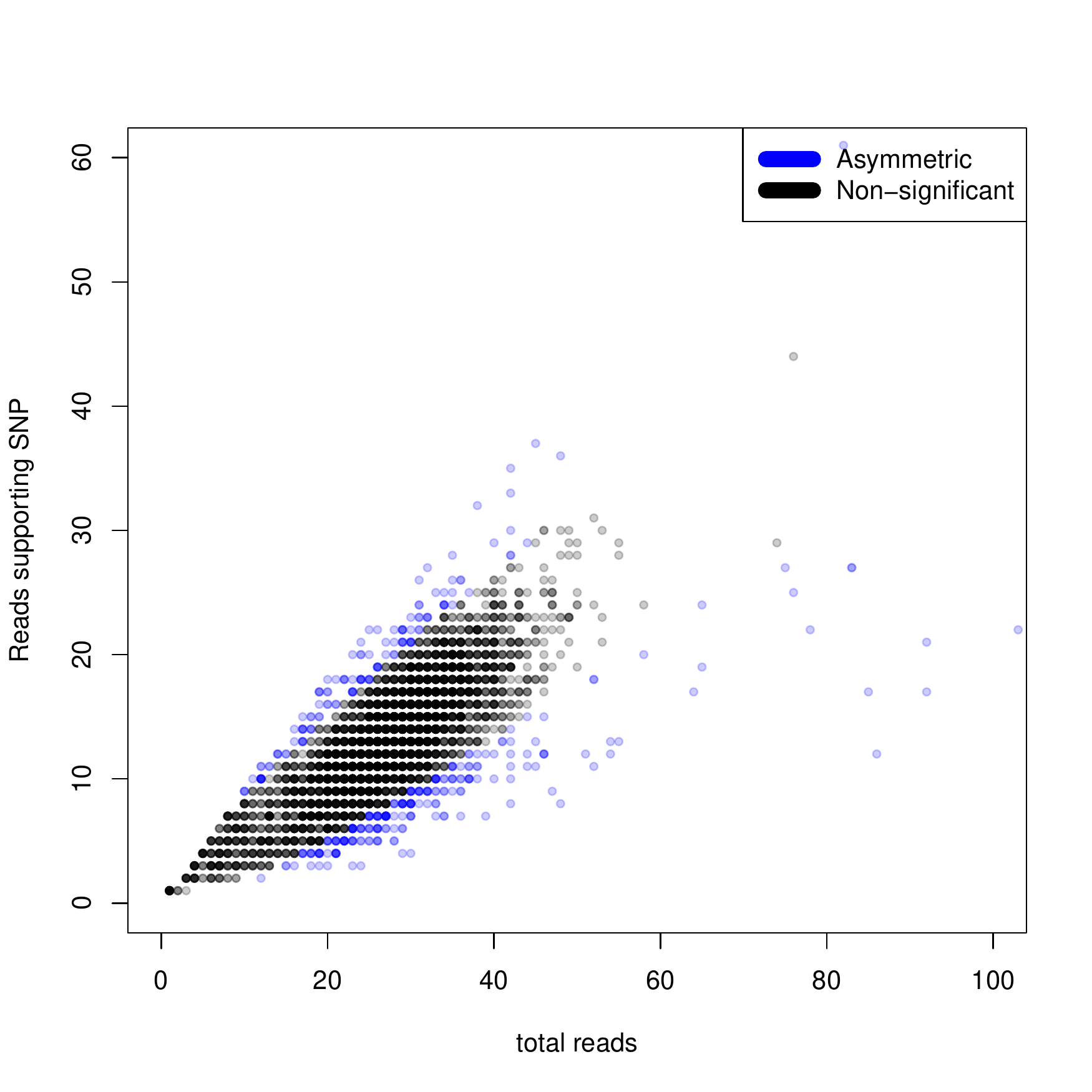}
 \caption{\label{AsymMaya} Maya}
      
    \end{subfigure}
\caption{  \label{AsymFig} Asymmetric SNPs out of 5000 representative SNPs on chromosome 1. }
\end{figure}

\clearpage
 \begin{figure} 

 \center

\begin{subfigure}[b]{0.35\textwidth}
\includegraphics[scale=0.35]{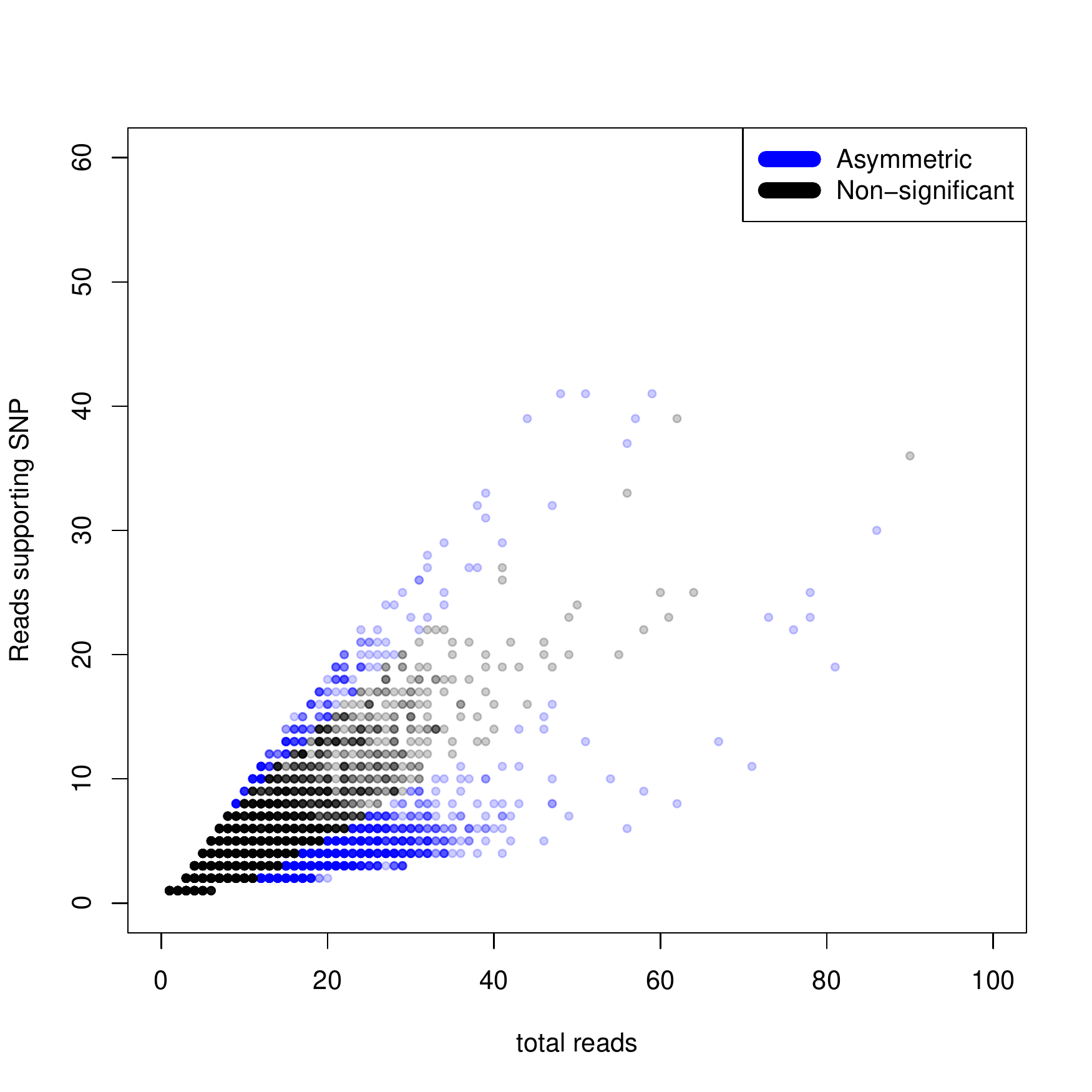}
 \caption{ M4 \label{AsymDamageM4}}
       
    \end{subfigure}
 \begin{subfigure}[b]{0.35\textwidth}
\includegraphics[scale=0.35]{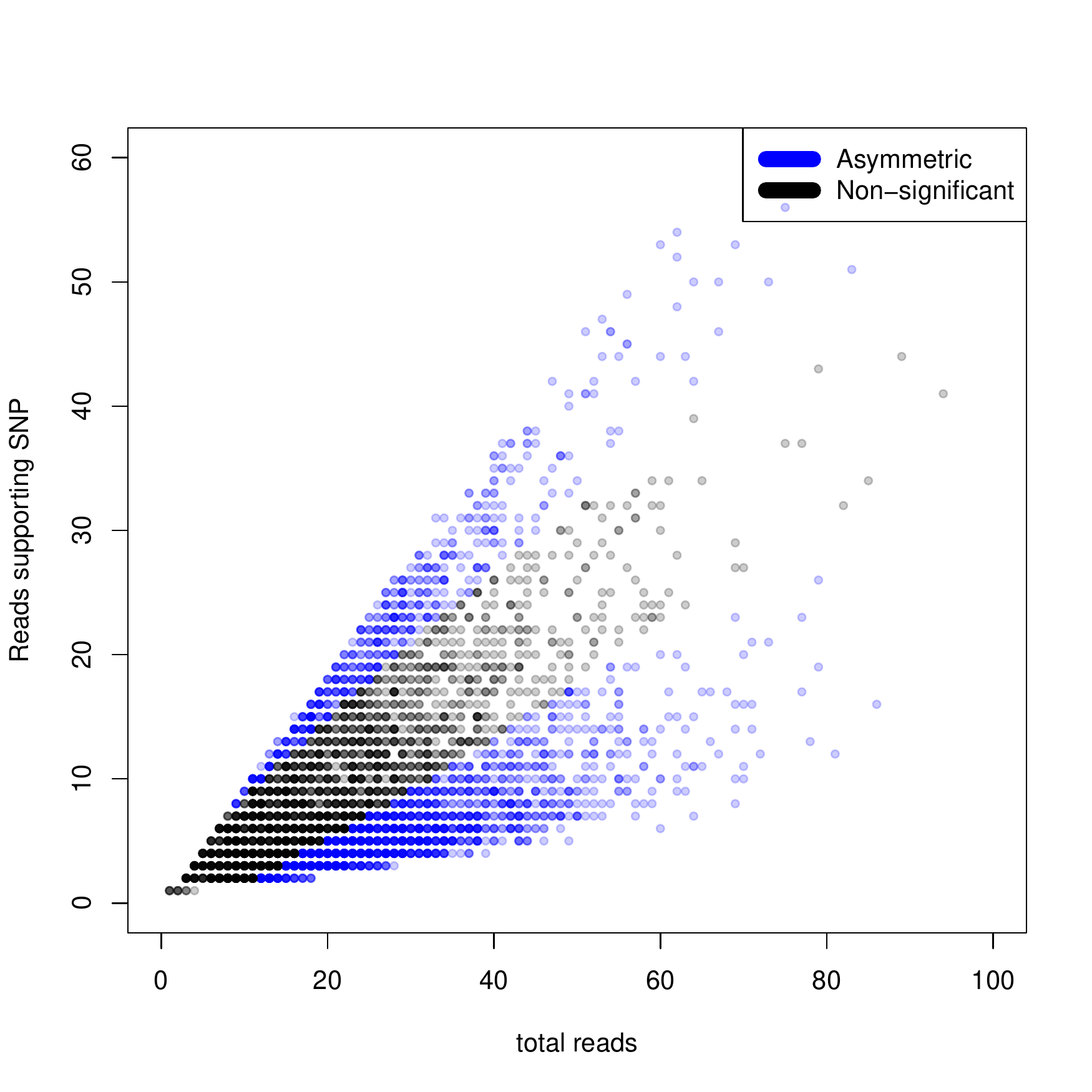}
 \caption{M25 \label{AsymDamageM25}} %
       
    \end{subfigure}

\caption{  \label{AsymDamageFig} Asymmetric SNPs out of 5000 representative SNPs on chromosome 1, excluding A/G and T/C mutations. }
\end{figure}

\end{document}